\documentclass[letterpaper,twoside,twocolumn,english,prl,intlimits,showpacs]{revtex4}
\usepackage{times}
\usepackage[T1]{fontenc}
\usepackage{amsmath}
\usepackage{graphicx}
\usepackage{amssymb}

\makeatletter

\usepackage{babel}
\makeatother
\begin{document}

\title{Front Propagation in Reaction-Superdiffusion Dynamics - Taming Lévy
Flights with Fluctuations}

\author{D.~Brockmann}

\author{L.~Hufnagel}

\affiliation{Max-Planck-Institut für Strömungsforschung, Bunsenstr. 10, 37073
Göttingen, Germany$^{\textrm{2}}$\\
and Kavli Institute for Theoretical Physics, University of California
Santa Barbara, CA 93106, USA}

\begin{abstract}
We investigate front propagation in a reacting particle system in
which particles perform scale-free random walks known as Lévy flights.
The system is described by a fractional generalization of a reaction-diffusion
equation. We focus on the effects of fluctuations caused by a finite
number of particles. We show that, inspite of superdiffusive particle
dispersion and contrary to mean field theoretical predictions, wave
fronts propagate at constant velocities, even for very large particle
numbers. We show that the asymptotic velocity scales with the particle
number and obtain the scaling exponent.
\end{abstract}


\maketitle

\newcommand{\differential}[1]{\textrm{d}#1}

\newcommand{\diff}[1]{\differential{#1}}

\newcommand{\dt}{\diff{t}}

\newcommand{\dx}{\diff{\zvec{x}}}

\newcommand{\dy}{\diff{y}}

\newcommand{\pd}[1]{\partial_{#1}\, }

\newcommand{\pdt}{\pd{t}}

\newcommand{\fraclp}[2]{\Delta_{#2}^{#1}}

\newcommand{\fraclpmu}[1]{\fraclp{\mu/2}{#1}}

\newcommand{\laplace}[1]{\fraclp{}{#1}}

\newcommand{\laplacex}{\laplace{x}}

\newcommand{\op}[1]{\mathcal{#1}}

One of the fundamental processes involved in non-equilibrium pattern
formation is the spatial propagation of interfaces or fronts. Front
propagation usually emerges when a local reaction dynamics interplays
with diffusion in space of the reacting agents and has been observed
in a wide range of physical, chemical and biological systems~\cite{cross_00851:1993}.
Prominent examples are patterns of bacterial colonies~\cite{goldi_00510:1998},
spreading phenomena in population genetics~\cite{fishe_00355:1937,arons_00033:1978}
and interface dynamics in superconducters~\cite{barto_04442:1996}.
One of the most prominent models which displays propagating fronts
is the Fisher-Kolmogorov-Petrovsky-Piscounov (FKPP) equation for the
spatial concentration $u(x,t)$ of a reacting agent, \begin{equation}
\pdt u=\gamma\, u\,(1-u)+\laplace{}u,\label{eq:fisher}\end{equation}
where diffusive motion of the reacting agents is assumed. However,
this assumption cannot be justified for a number of systems. In fact,
superdiffusive dispersion in space has been observed in a wide range
of physical and biological systems, e.g. intermittent chaotic systems~\cite{geise_00616:1985},
bacterial motion~\cite{levan_00237:1997}, and foraging patterns
of albatrosses~\cite{viswa_00413:1996}.

Superdiffusive stochastic motion is usually characterized by a lack
of scale in the microscopic single step distribution. One of the most
successful theoretical concepts devised for the understanding of superdiffusion
is a class of random walks known as Lévy flights~\cite{shles:1995}.
A Lévy flight consists of random single steps $\Delta x$ which are
drawn from an inverse power-law pdf (probability density function)
$p(\Delta x)\sim|\Delta x|^{-(1+\mu)}$ characterized by a Lévy exponent
$0<\mu<2$. Due to the heavy tail, the variance in step size is divergent,
the process lacks a spatial scale and the position $X(t)$ of a Lévy
flight scales heuristically with time $t$ as $X(t)\sim t^{1/\mu}$.
The associated diffusion equations contain fractional generalizations
of ordinary derivatives~\cite{metze_00001:2000,sokol_00048:2002}.
These fractional Fokker-Planck equations can exhibit behaviors strikingly
different from ordinary ones~\cite{brock_170601:2003} and have found
wide application in physics, e.g. protein motion on folded hetero-polymers~\cite{brock_048303:2003}
and the dynamics of modern epidemics~\cite{SARSDEAMON:2004,SIRDEAMON:2004}.

In two recent studies wave front dynamics was shown to be drastically
different from ordinary reaction-diffusion dynamics when the reacting
agents move superdiffusively~\cite{manci_00532:2002,del_ca_018302:2003}.
The authors considered fractional generalizations of Eq.~(\ref{eq:fisher})
and showed that the spatio-temporal shape $u(x,t)$ of the leading
egde of a propagating front has a power-law tail along the spatial
coordinate and accelerates exponentially in time as opposed to the
constant velocity and exponential decay in space exhibited by ordinarily
diffusive system. The predictions made by mean field theory thus indicate
that scale-free, superdiffusive dispersion of the reacting agents
excludes constant velocity wave fronts and induces an entirely different
spatio-temporal behavior.

However, as has been shown in a number of recent studies, the effect
of fluctuations can be rather profound in these systems~\cite{peche_03893:1999,brune_00269:2001,doeri_00243:2003}.
For instance, fluctuations can destabilize homogeneous wave fronts
in two-dimensional systems~\cite{kessl_00556:1998}. A finite albeit
large number $N$ of particles or reacting agents leads to a multiplicative
noise term in the reaction-diffusion equations and although the variance
of the noise is of order $1/N$, corrections to macroscopic quantities
such as the front speed $v$ scale as $(\log N)^{-2}$, a significant
correction to the mean field approximation even for very large values
of $N$. Brunet and Derrida~\cite{brune_02597:1997,kessle_00107:1998}
extended mean field dynamics by an effective cutoff parameter $\varepsilon$
for the concentration of particles below which no reaction and hence
no exponential growth of the leading edge of a front is possible.
Despite the fact that a rigorous equivalence with multiplicative noise
is still lacking, the effective cutoff approach is very intuitive
and in remarkable agreement with simulations of the full probabilistic
dynamics.

Here, we focus on the effect of fluctuations on reaction-superdiffusion
kinetics. We show that for arbitrarily small fluctuations (i.e. arbitrarily
large particle numbers $N$), wave fronts propagate asymptotically
at constant velocities. Furthermore, we show that as soon as fluctuations
enter the description the algebraic tail along the spatial coordinate
of the leading edge disappears and is replaced by an exponential decay.
Thus, despite the fact that reacting agents move superdiffusively
in space, the wave front patterns are qualitatively the same as in
the ordinary diffusion case. We show that a front speed $v$ is selected
after a transient time and that $v$ scales with particle number as
$v\sim N^{1/\mu}$ for Lévy exponents $\mu<2$. The results reported
here are rather counterintuitive, deviate strongly from the predictions
cast by mean field theory, and indicate that fluctuations affect reactions-superdiffusion
systems severely and may by no means be neglected. 

\begin{figure}
\includegraphics[%
  width=1.0\columnwidth]{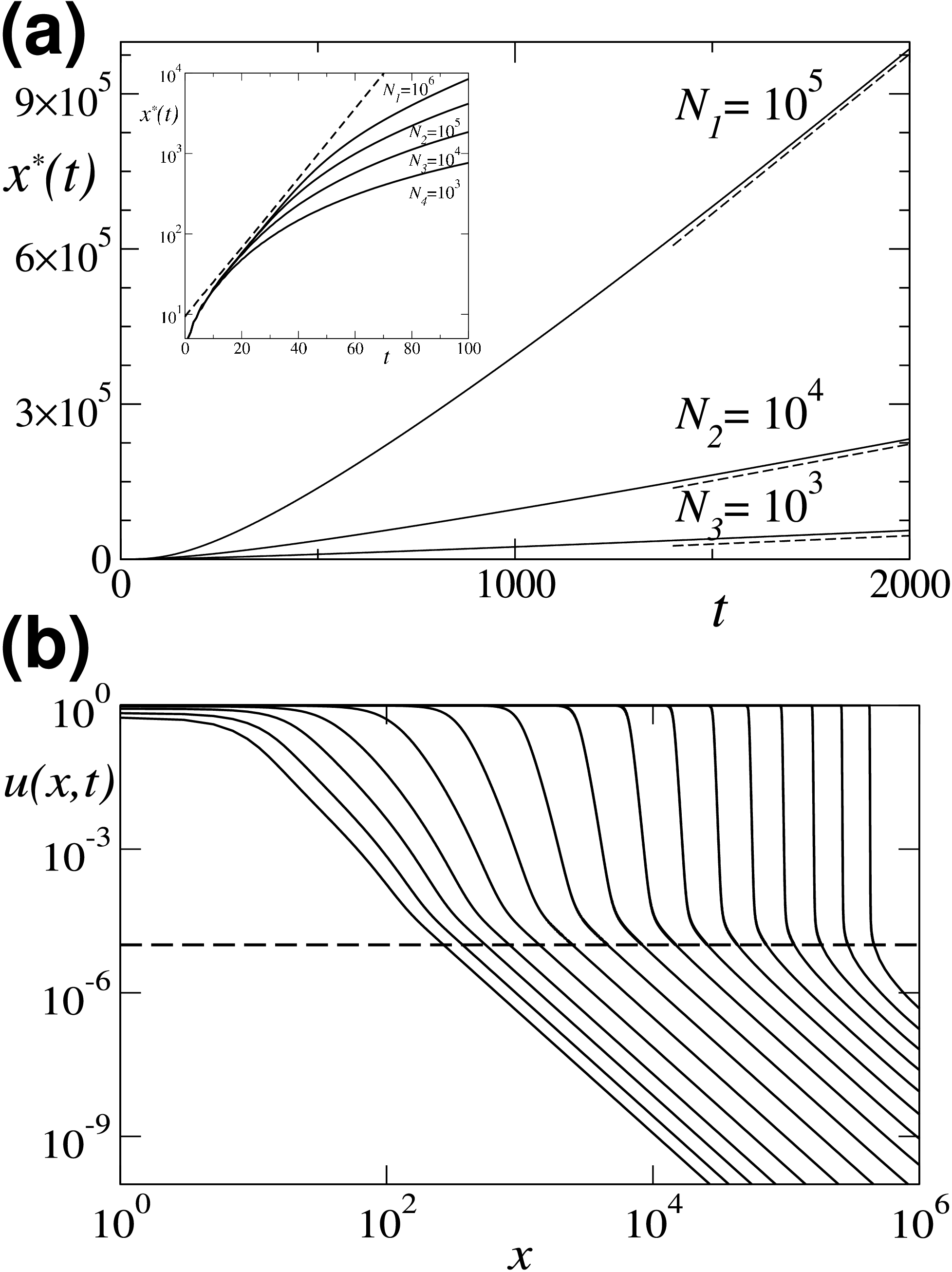}

\caption{\label{cap:fig1} Propagation of wave fronts in superdiffusive systems
with effective cutoff. (a) Location of the wave front $x^{*}(t)$
as a function of time $t$ for different particle numbers $N$ (solid
lines) for $\mu=1.5$ and $\gamma=0.1$ in Eq.~(\ref{eq:schwelle1}).
Following initial transients the velocity $v=\diff{x}^{*}(t)/\dt$
is constant (dashed lines) and increases with $N$. The inset depicts
a magnification of the initial phase. The wave front accelerates exponentially
but deviates from the mean-field dynamics ($\varepsilon=0$, dashed
line) after the transient period. (b) Shape of the wave front at exponentially
increasing time steps $t=1.5^{m}$ with $m=5,6,...,20$ for $N=10^{4}$.
After a transient phase the shape remains unaltered, decays sharply
for $1/N<u(x,t)<1$ and follows a power law $u(x,t)\sim x^{_{-(1+\mu)}}$
for large $x$. The dashed line indicates the effective cutoff $\varepsilon=10^{-4}$.
The spatial extent in the numerical integration was $L=2^{23}\approx8.38\times10^{6}$.}
\end{figure}
We begin with a simple two particle type ($A,\, B$) reaction scheme,\begin{gather}
A_{x}+B_{x}\xrightarrow{k_{1}}2A_{x},\qquad A_{x}+B_{x}\xrightarrow{k_{2}}2B_{x}\label{eq:reaction}\\
A_{x},B_{x}\xrightarrow{f(|x-y|)}A_{y},\, B_{y}.\label{eq:dispersion}\end{gather}
Particles of type $A$ and $B$ either react to produce two particles
of type $A$ or two particles of type $B$ at rate $k_{1}$ and $k_{2}$,
respectively. Furthermore, particles of both types may jump from position
$x$ to position $y$ with a probability density rate $f(|x-y|)$
which we assume to be a decreasing function of distance $|x-y|$.
The dynamic stochastic variables are the particle numbers $n_{A}(x,t)$
and $n_{B}(x,t)$ of particles in an volume of size $\Omega$ around
$x$ of type $A$ and $B$, respectively. The volume $\Omega$ is
assumed to be large enough to contain a large number of particles
but small compared to the spatial extend of the system. The total
number of particles in $\Omega$ around $x$ is given by $N(x,t)=n_{A}(x,t)+n_{B}(x,t)$.
Without spatial dispersion of particles the local dynamics is governed
by the master equation \begin{equation}
\pdt p(n,t)=\sum_{m}w(n|m)p(m,t)-w(m|n)p(n,t)\label{eq:master}\end{equation}
for the probability $p(n,t)$ of finding a number $n=n_{A}$ of particles
of type $A$ at a location $x$ with initial condition $p(n,0)=\delta(n-n_{0})$
and the rate\begin{equation}
w(n|m)=\frac{k_{1}}{\Omega}m(N\!-\! m)\delta_{n,m+1}+\frac{k_{2}}{\Omega}m(N\!-\! m)\delta_{n,m-1}.\label{eq:rate}\end{equation}
The dynamics of the expectation value of $\left\langle n(t)\right\rangle $
is governed by\begin{equation}
\pdt\left\langle n(t)\right\rangle =\frac{\gamma}{\Omega}\left\langle n(t)\,(N-n(t))\right\rangle ,\label{eq:localexpect}\end{equation}
where $\gamma=k_{1}-k_{2}$. Note that $\left\langle n(t)\right\rangle $
is continuous in $[0,N]$. In a spatially extended system the number
of particles is a function of position, i.e. $n=n(x,t)$. Apart from
normalization, $\left\langle n(x,t)\right\rangle $ may be interpreted
as the probability of finding an $A$ particle in the volume $\Omega$.
Dispersion contributes to the change of particles $\partial_{t}\left\langle n(x)\right\rangle $
according to\begin{equation}
\pdt\left\langle n(x,t)\right\rangle =\int\dy\, f(|x-y|)\left[\left\langle n(y,t)\right\rangle -\left\langle n(x,t)\right\rangle \right],\label{eq:dispersexpect}\end{equation}
incorporating the probability density rate (Eq.~(\ref{eq:dispersion}))
of jumping from $y$ to $x$. Eq.~(\ref{eq:dispersexpect}) defines
the operator $\op{L}$ acting on the field $\left\langle n(x,t)\right\rangle $.
Denoting the spatial density of particles by $u(x,t)=n(x,t)/\Omega$
and combining Eq.~(\ref{eq:localexpect}) with~(\ref{eq:dispersexpect})
yields\begin{equation}
\pdt\left\langle u\right\rangle =\gamma\left\langle u\,(1-u)\right\rangle +\op{L}\left\langle u\right\rangle ,\label{eq:pupsi}\end{equation}
where we have set without loss of generally the maximum density $N/\Omega$
to unity. 

When particles perform ordinary random walks, i.e. if the jump rate
$f(|x-y|)$ is equipped with a length scale $\sigma$, the operator
$\op{L}$ can be approximated by the ordinary Laplacian $\laplace{}$
on scales larger than $\sigma$. However, this description is no longer
valid if $f(x)\sim|x|^{-(1+\mu)}$ with $0<\mu<2$. In this case individual
jumps lack a scale, particle perform Lévy flights, and $\op{L}$ is
proportional to a non-local singular integral operator, \begin{equation}
\fraclpmu{}u(x,t)=C_{\mu}\int\diff{y}\frac{\left[u(y,t)-u(x,t)\right]}{|x-y|^{1+\mu}}.\label{eq:fractionallaplace}\end{equation}
where $C_{\mu}$ is a constant~\cite{brock_00409:2002}. In Fourier
space the operator $\fraclpmu{}$ is equivalent to a multiplication
by $|k|^{\mu}$, generalizing the well known $k^{2}$ factor corresponding
to the ordinary Laplacian which is why $\fraclpmu{}$ is frequently
referred to as a fractional Laplacian.

Since $\left\langle u^{2}\right\rangle >\left\langle u\right\rangle ^{2}$
one cannot computed the dynamics of $\left\langle n(x,t)\right\rangle $
from Eq.~(\ref{eq:pupsi}) alone. The crudest approximation is the
mean field approach in which fluctuations are ignored ($\left\langle u^{2}\right\rangle \approx\left\langle u\right\rangle ^{2}$and
thus $u=\left\langle u\right\rangle $) and Eq.~(\ref{eq:pupsi})
reads\begin{equation}
\pdt u=\gamma\, u\,(1-u)+\fraclpmu{}u,\label{eq:genmeanfield}\end{equation}
which is the Fisher-equation~(\ref{eq:fisher}) for $\mu=2$ and
its superdiffusive mean field generalization for $\mu<2$.

The next better step is to account for fluctuations ($N\gg1$) which
leads to an additional multiplicative white noise term of variance
$u(1-u)/N$ in Eq.~(\ref{eq:genmeanfield}). Alternatively one may
incorporate the finite particle number as an effective cutoff $\varepsilon\sim1/N$
in the logistic growth term retaining a deterministic equation, i.e.\begin{equation}
\pdt u=\gamma\, u\,(1-u)\Theta(u-\varepsilon)+\fraclpmu{}u,\label{eq:schwelle1}\end{equation}
 in which $\Theta$ is the Heaviside-function. Qualitatively, this
accounts for the fact that no growth on average can occur if the particle
concentrations $u$ is less than one particle per unit volume. This
approach has been applied successfully in the ordinary diffusion scenario~\cite{brune_02597:1997}.
In Eq.~(\ref{eq:schwelle1}), mean field theory implies $\varepsilon=0$. 

\begin{figure}
\includegraphics[%
  width=1.0\columnwidth]{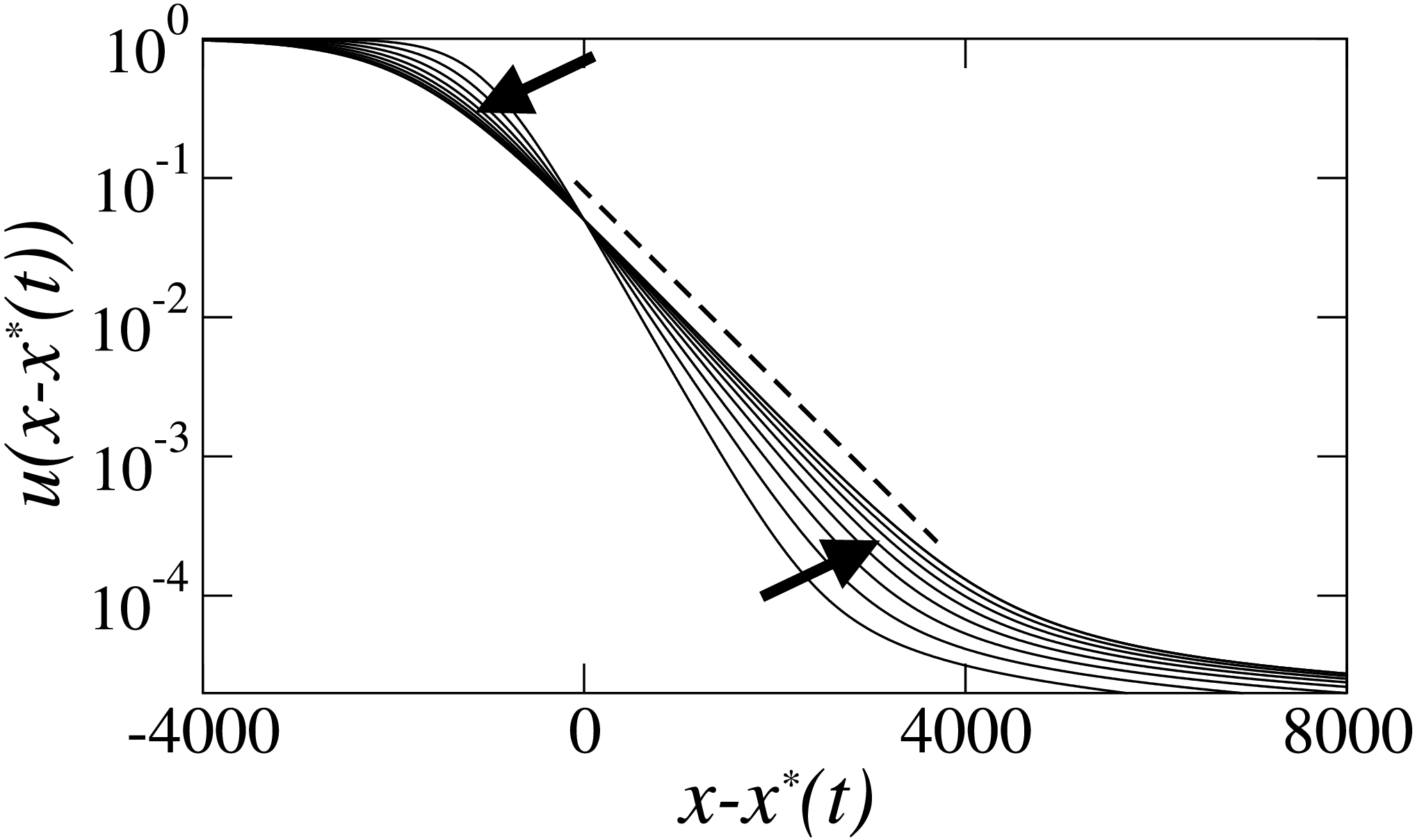}

\caption{\label{cap:fig2}Scaling behavior of the fronts profiles at exponentially
increasing times $t$ as in Fig.~\ref{cap:fig1} in the reference
frame of the front. Arrows indicate the temporal direction. The curves
approach a steady state characterized by an exponential decay in space,
$\exp-\lambda x$. The position of the from $x^{*}(t)$ is given by
$u(x^{*}(t),t)=0.05$ . Other parameters are as in Fig.~\ref{cap:fig1}.}
\end{figure}
 In the following we investigate front dynamics of the system defined
by Eq.~(\ref{eq:schwelle1}) and compare its properties to the mean
field dynamics ($\varepsilon=0$) , the limiting case of ordinary
diffusion ($\mu=2$) and the dynamics of the full probabilistic dynamics
(Eqs.~(\ref{eq:reaction}) and~(\ref{eq:dispersion})).

Fig.~\ref{cap:fig1} depicts front dynamics obtained from numerical
integration of~(\ref{eq:schwelle1}) for a concentration $u(x,t)$
initially peaked at the origin. For various particle numbers $N$
the velocity and shape of the front were computed. Fig.~\ref{cap:fig1}(a)
shows that even for very large particle numbers wave fronts move asymptotically
at constant speeds, a remarkable difference from the mean field limit
(dashed line in the inset) which predicts exponentially accelerating
front. Fig.~\ref{cap:fig1}(b) depicts snapshots of the wavefronts
on a double-logarithmic scale. After a transient phase, the shape
of the front approaches a steady state with a sharply decreasing boundary
at intermediate values for the concentration and an algebraic tail
for large $x$ with concentrations $u(x,t)$ below the cutoff $\varepsilon$.
Fig.~\ref{cap:fig2} displays the steady state front shape on a semilogarithmic
scale. As opposed to the algebraic tail predicted by mean field theory,
the boundary is an exponential function of relative position $z=x-vt$,
i.e. $u(z)\sim\exp-\lambda z$, characterized by a spatial scale $\lambda$.
In summary, the characteristic spatio-temporal wave front solution
of Eq.~(\ref{eq:schwelle1}) for large times is given by $u(x,t)\sim\exp\left[-\lambda(x-vt)\right],$
for $1/N<u<1$ followed by a power-law tail $u(x,t)\sim(x-vt)^{-(1+\mu)}$
for $u<1/N$. The decay parameter $\lambda$ and the velocity $v$
depend on the particle number $N$. Qualitatively, the this dependence
can be determined in the moving reference frame under the assumption
that $u(x,t)=u(x-vt)=u(z)$. Figs.~\ref{cap:fig1}(b) and~\ref{cap:fig2}
suggest that $u(z)=u_{1}(z)=\exp-\lambda z$ for $z<z^{\star}$ and
$\varepsilon<u\ll1$ and $u(z)=u_{2}(z)=A/z^{1+\mu}$ for $z>z^{\star}$
($u<\varepsilon$) where $z^{\star}$ marks the crossover between
exponential and algebraic decay. Inserted into Eq.~(\ref{eq:schwelle1})
yields\begin{eqnarray}
v\lambda u_{1} & \approx & u_{1}+\fraclpmu{}u\qquad z<z^{\star}\label{eq:katze}\\
-v\, u_{2}^{\prime} & = & \fraclpmu{}u\qquad z>z^{\star}.\label{eq:hund}\end{eqnarray}
Near the crossover the values of the dispersion in both equations
are approximately the same, i.e. $(\fraclpmu{}u)(z^{\star}-\Delta z)\approx(\fraclpmu{}u)(z^{\star}+\Delta z)$
and both equations can be combined to $v\lambda u_{1}\approx u_{1}-v\, u_{2}^{\prime}$.
With $u_{2}^{\prime}(z^{\star})=-(1+\mu)u_{2}(z^{\star})/z^{\star}$
and $u_{2}(z^{\star})\approx u_{1}(z^{\star})$ we obtain $v\lambda\approx1-v(1+\mu)/z^{\star}$.
Since $z^{\star}\gg1$ the second term can be neglected and we find\begin{equation}
v\sim\lambda^{-1}.\label{eq:faulambda}\end{equation}
The dependence of $\lambda$ on $N$ can be obtained by the short
time dynamics. Consider an initial condition $u(x,0)=\exp(-\lambda x)$
on the half-line $x\ge0$. After a short time $\Delta t\ll1$ the
wave front is approximately given by $u(x,\Delta t)\approx\exp-\lambda x+\Delta t/\lambda x^{1+\mu}.$
The crossover $\overline{x}$ can be defined as the point at which
both terms are of the same order of magnitude, implicitly provided
by $\lambda\exp-\lambda\overline{x}=\Delta t/\overline{x}^{1+\mu}$.
Since $\lambda\ll1$ one can approximately solve for the crossover,
$\overline{x}\approx\ln(\lambda^{\mu}\Delta t)/\lambda$. In order
for the exponential to remain invariant under the dynamics, $\lambda$
must be chosen such that the crossover coincides with the effective
cutoff, i.e. $u(\overline{x},\Delta t)=\varepsilon$ which implies
the scaling relation $\lambda\propto\varepsilon^{1/\mu}$. With Eq.~(\ref{eq:faulambda})
one obtains the scaling law

\begin{equation}
v\propto N^{1/\mu}\label{eq:vscaling}\end{equation}
for the velocity $v$.

\begin{figure}
\includegraphics[%
  width=1.0\columnwidth]{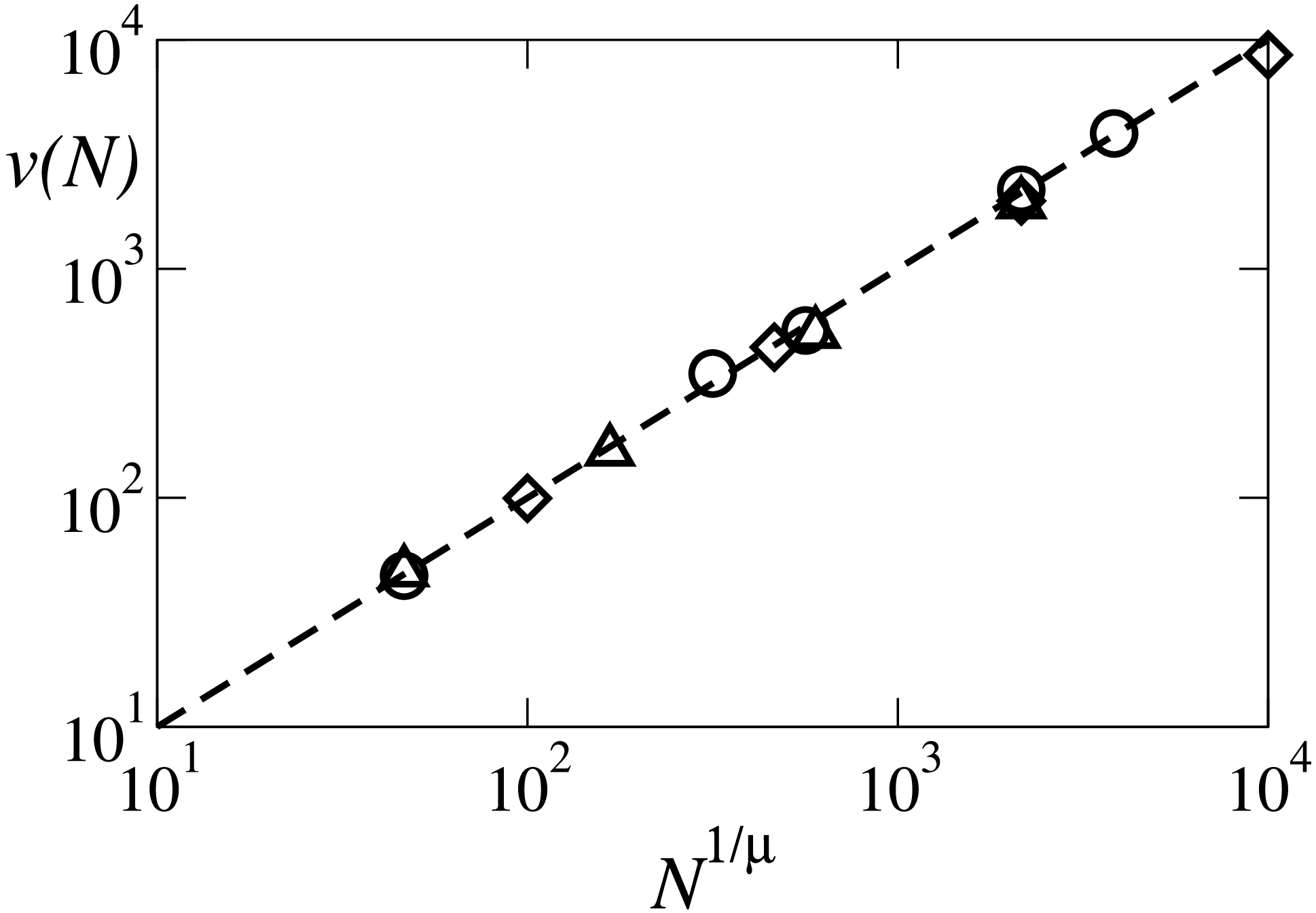}

\caption{\label{cap:fig3}Asymptotic front velocity $v$ as a function of
the particle number $N$ for different Lévy exponents $\mu=1.2,\,1.5$
and $1.8$ (circles, diamonds and triangles, respectively). The dashed
line indicates the scaling $v(N)\sim N^{1/\mu}$.}
\end{figure}
 Fig.~\ref{cap:fig3} shows the front velocity $v$ as a function
of particle number $N$ obtained by numerical integration of the dynamics
(Eq.~(\ref{eq:schwelle1})). The numerics agree well with the scaling
law~(\ref{eq:vscaling}) over several orders of magnitude and several
choices of the Lévy exponent $\mu$.

\begin{figure}
\includegraphics[%
  width=1.0\columnwidth]{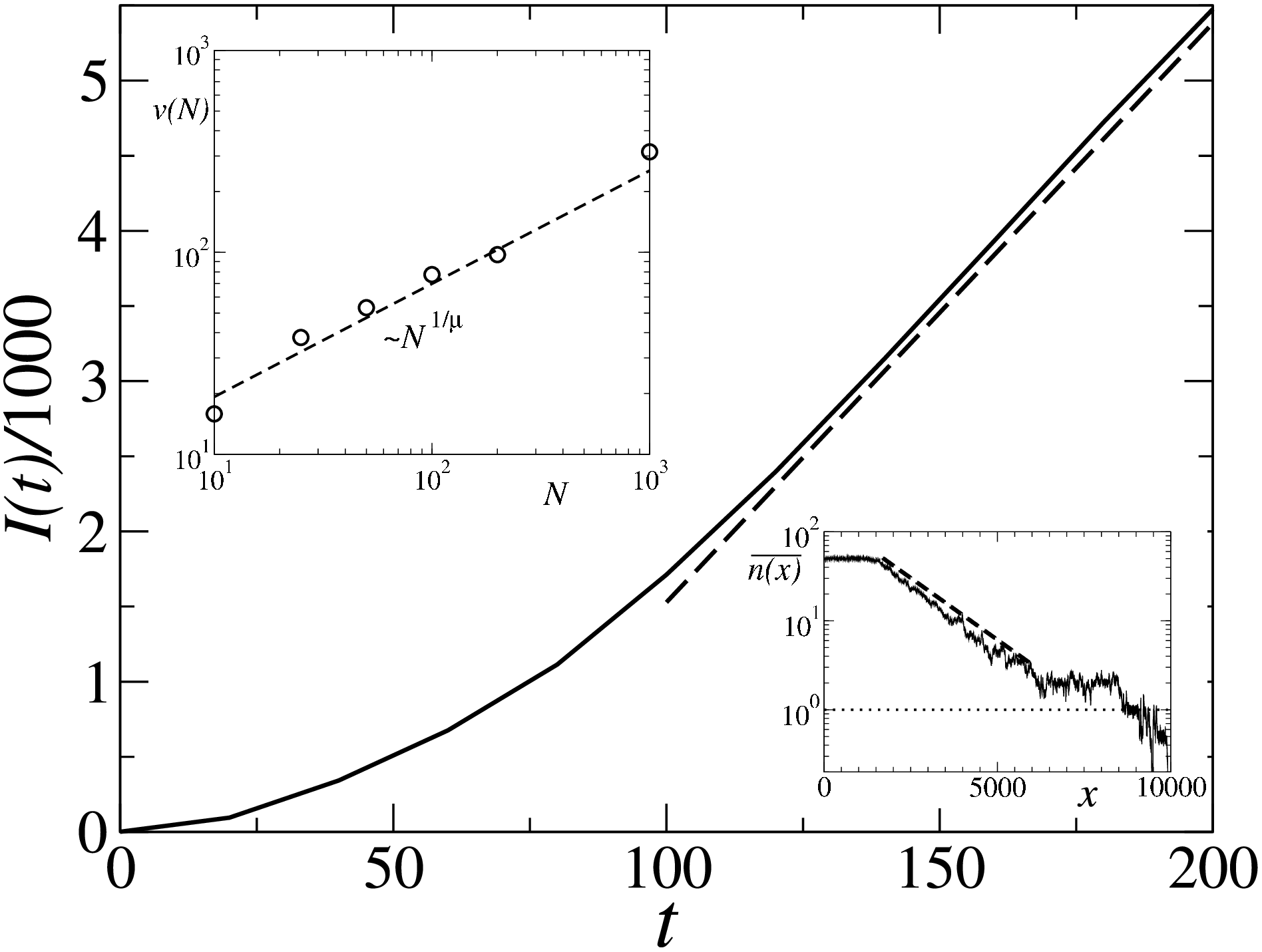}

\caption{\label{cap:fig4}Asymptotic front velocity in the full stochastic
model. The solid line depicts the total mass $I(t)$ type $A$ particles
as a function of time. The system size is $L=10^{4}$, the total number
of particles in the system $2.5\times10^{4}$. Other parameters are
$\gamma=1$ and $\mu=1.8$. A constant front velocity (constant $\diff{I}(t)/\dt$)
is attained asymptotically (dashed line). The upper inset depicts
the front velocity $v(N)$ as a function of particles per site for
$N=10,\,25,\,50,\,100,\,200$ and $1000$ and the scaling law~(\ref{eq:vscaling})
(dashed line). The lower inset depicts an average $\overline{n(x)}$
of $A$ particles at position $x$ at an intermediate time $t$. The
average was computed from a sample of $100$ realizations of the process.
The dashed line indicates an exponential decay, the dotted line the
average concentration of one particle per site.}
\end{figure}
 The front propagation characteristics of the macroscopic description
in terms of the reaction-superdiffusion equation with effective cutoff
(Eq.~(\ref{eq:schwelle1})) coincides with our simulations of the
microscopic dynamics defined by Eqs.~(\ref{eq:reaction}) and~(\ref{eq:dispersion}).
The simulation results are summarized in Fig.~\ref{cap:fig4}. As
in Fig.~\ref{cap:fig1}(a) the front velocity $v$ is constant after
a transient phase, the velocity obeys the scaling law~(\ref{eq:vscaling})
and the wave front decays exponentially in space. 

We are convinced that our results are of major importance for the
understanding of front propagation in pattern forming systems in which
the reactive agents defy the rules of ordinary diffusion. We have
shown that constant velocity fronts are typical for pulled front dynamics,
contrary to what is expected from mean field approximations and we
believe that our results will contribute to the understanding of more
complex pattern forming systems such as the geographic spread of human
epidemics~\cite{SIRDEAMON:2004}.

\begin{acknowledgments}
This research was supported in part by the National Science Foundation
under grant No.~PHY99-07949. Discussions with T. Geisel, L. Sander,
W. Noyes, V. Hardapple and M. Boone Jr. are greatfully acknowledged.
\end{acknowledgments}
\bibliographystyle{apsrev}
\bibliography{./simon}

\end{document}